# Full Temperature-Dependent Potential and Anharmonicity in Metallic Hydrogen: Colossal NQE and the Consequences

Hua Y. Geng[1,2*]

[1]*National Key Laboratory of Shock Wave and Detonation Physics, Institute of Fluid Physics, CAEP; P.O.Box 919-102 Mianyang, Sichuan, P. R. China, 621900*

[2]*HEDPS, Center for Applied Physics and Technology, and College of Engineering, Peking University, Beijing 100871, People's Republic of China*

Corresponding author:
Hua Y. Geng
National Key Laboratory of Shock Wave and Detonation Physics, Institute of Fluid Physics, CAEP; P.O.Box 919-102 Mianyang, Sichuan, P. R. China, 621900
Email address: s102genghy@caep.cn

## Novelty and Significance

(1) Widely used TDEP method for anharmonic phonon is generalized to the full potential case (*i.e.*, FTDP), which naturally resolves the notorious low-temperature problem of the former.
(2) Very strong nuclear quantum effect (NQE) in metallic hydrogen is discovered via FTDP and direct path integral molecular dynamics (PIMD) calculations.
(3) Possible breakdown of phonon picture in solid metallic hydrogen is indicated for the first time, which implying novel lattice dynamical phenomena unknown so far might exist.
(4) FTDP+PIMD is established as a helpful method to probe colossal NQE and possible failure of phonon picture in quantum solid.
(5) A general ionic dynamics (GID) formalism for quantum lattice dynamics beyond phonon is proposed, with the basic features are illustrated.

* Email: s102genghy@caep.cn

# ABSTRACT

The temperature-dependent effective potential (TDEP) method for anharmonic phonon dispersion is generalized to the full potential case by combining with path integral formalism. This extension naturally resolves the intrinsic difficulty in the original TDEP at low temperature. The new method is applied to solid metallic hydrogen at high pressure. A colossal nuclear quantum effect (NQE) and subsequent anharmonicity are discovered, which not only leads to unexpectedly large drift of protons, but also slows down the convergence rate substantially when computing the phonon dispersions. By employing direct ab initio path integral molecular dynamics simulations as the benchmark, a possible breakdown of phonon picture in metallic hydrogen due to colossal NQE is indicated, implying novel lattice dynamical phenomena might exist. Inspired by this observation, a general theoretical formalism for quantum lattice dynamics beyond phonon is sketched, with the main features being discussed.

## I. INTRODUCTION

Condensed matter is dictated by quantum mechanics of electrons and ions, with the former gives rise to local or non-local orbitals and subsequent chemical bonding, which add up to inter-atomic interactions and govern the movement of the latter(1). At low temperatures, ions vibrate around their respective equilibrium positions collectively that is determined by the potential energy surface (PES). When in a crystalline phase, quantization of this motion leads to phonon as quasi-particles(2). By far, the state-of-the-art theory for lattice dynamics of solid is exclusively built on phonons, with high-order contributions are treated as perturbation and interpreted as scattering between phonons(1)(2). In this formalism, non-scattered phonons have infinite lifetime (*i.e.*, undamped) and correspond to a *quadratic* (or harmonic) PES(3)(4). Any deviation from this harmonic approximation (HA) is regarded as anharmonic, and must have a non-quadratic PES.

Increasing evidences have demonstrated that strong anharmonicity could lead to imaginary modes in HA phonons, making perturbation on top of it impossible. The theory of self-consistent harmonic approximation (SCHA)(5)(6) evaded this problem by using an *effective* harmonic model to approximate the true system, with the aid of cumulant expansion and the principle of minimum free energy for the model system(7). The lowest order solutions of SCHA are undamped. They do not correspond to bare phonons, but are instead dressed by contributions from even-order derivatives of the PES. Namely, they are quasi-particles already been *renormalized* by anharmonicity over a finite region around the equilibrium position given by the respective SCHA modes(7). In the following we will refer them as *quasi-phonon* for brevity (to distinguish them from the bare phonon of HA). Perturbations on top of quasi-phonon can be applied if the corresponding expansion converges rapidly enough. Unfortunately, no rigorous implementation of this level of lattice dynamics has been reported yet, at least to the best of our knowledge. Straightforward extension of this approach remains in the realm of phonon picture as represented by quasi-phonons (and the scattering between them) if applied successfully.

In principle, it is possible that anharmonicity could be strong enough so that even quasi-phonon becomes invalid, and the lattice dynamics cannot be interpreted as the scattering of freely moving quasi-particles any more, leading to a complete failure of the phonon picture.

To probe this extreme possibility, in this work, the temperature-dependent effective potential (TDEP) method for anharmonic phonon is generalized to a new level so that has the capability of exploring the full physical PES by combining with path integral molecular dynamics (PIMD) simulations. This full temperature-dependent potential (FTDP) extension is then applied to high pressure metallic hydrogen (MH). The results not only uncover colossal NQE and strong anharmonicity, but also indicate that in MH the anharmonicity is so strong that the phonon picture might become inadequate for the lattice dynamics.

In next section, the theoretical formalism of TDEP and FTDP are presented, as well as their implementation and the combination of the latter with PIMD. Validation of the method, together with the application to MH, is provided in Sec. III. Further discussions and the extension of the lattice dynamics beyond phonon are presented in Sec. IV, with Sec. V summarizes the main findings.

## II. THEORY AND METHODOLOGY

### A. Theory Formalism

#### A1. TDEP

Traditional lattice dynamics is constructed on the perturbation treatment to the potential energy of a crystal with respect to small displacements on the PES. The Born-von Karman expansion of the PES gives an infinite series of

$$V\text{-}V_0=\frac{1}{2}\sum_{m_1\alpha_1,m_2\alpha_2}\sum_{\mu_1,\mu_2}\Phi^{\mu_1,\mu_2}_{m_1\alpha_1 m_2\alpha_2}\times u^{\mu_1}_{m_1\alpha_1}u^{\mu_2}_{m_2\alpha_2}$$

$$+\sum_{N=3}^{\infty}\frac{1}{N!}\sum_{m_1\alpha_1,\cdots,m_N\alpha_N}\sum_{\mu_1,\cdots,\mu_N}\Phi^{\mu_1,\cdots,\mu_N}_{m_1\alpha_1,\cdots,m_N\alpha_N}\times u^{\mu_1}_{m_1\alpha_1}\cdots u^{\mu_N}_{m_N\alpha_N} \quad (1)$$

with the interatomic force constants (IFC) defined as

$$\Phi^{\mu_1,\cdots,\mu_N}_{m_1\alpha_1,\cdots,m_N\alpha_N}=\left(\frac{\partial^N \mathrm{V}}{\partial u^{\mu_1}_{m_1\alpha_1}\cdots\partial u^{\mu_N}_{m_N\alpha_N}}\right)_0 \quad (2)$$

Here V denotes the potential energy, $u^{\mu}_{m\alpha}$ is the displacement of particle $\mu$ in the unit cell $m$ along the Cartesian component $\alpha$. The subscript 0 indicates the quantity being evaluated at the equilibrium position. The first-order terms in the expansion of Eq.(1) are null according to the stationary condition at the equilibrium. The assumption of small relative amplitude of the displacement $u$ against the inter-nuclear separations allows one to truncate Eq.(1) to retain only the second-order derivatives. This gives rise to the harmonic or quasi-harmonic phonons (QHA). On the other hand, when the high-order derivatives are small enough, perturbation theory can be applied on top of QHA to take anharmonic corrections into account. Unfortunately, in some rare-gas crystals, the convergence of Eq.(1) is very slow, and is even nonexistent for helium.

Self-consistent phonon (SCP)[(3)(4)(5)(6)(7)] is a non-perturbative method which is adequate for the lattice dynamics of anharmonic solid such as helium. In SCP, a model of phonon dynamics with adjustable parameters is constructed and then optimized to approximate the low-lying excitation spectra of the true system as close as possible. The physics requirement is satisfied by the variational principle of lattice free energy (in which $\beta^{-1}=k_{\mathrm{B}}T$)

$$F=-\beta^{-1}\ln\mathrm{Tr}[\exp(-\beta H)]\ \leq\ F^{\mathrm{SCP}}=-\beta^{-1}\ln\mathrm{Tr}[\exp(-\beta H^{\mathrm{SCP}})] \quad (3)$$

The potential energy in the model Hamiltonian $H^{\mathrm{SCP}}$ could be a quadratic form in the atomic displacements, with whose force constants $\tilde{\Phi}^{\mu_1,\mu_N}_{m_1\alpha_1,m_N\alpha_N}$ acting as freely adjustable parameters. It is evident that if $\tilde{\Phi}$ are frequency-independent, the introduced model phonons are free bosons; otherwise the associated phonons must be damped to a finite lifetime. The first-order approximation of SCP leads to the former case, in which only the even derivatives of Eq.(1) make contributions; whereas higher-order SCP requires frequency-dependent $\tilde{\Phi}$, so that both odd and even derivatives of Eq.(1) contributed to the model phonon damping rate and produced a frequency-shift(7).

It is obvious that in SCP the force constants $\tilde{\Phi}^{(2)}$ of a quadratic model were introduced to approximate the true PES $V$-$V_0$ in a sense of "mean-field" statistical average

$$H^{\mathrm{SCP}} = \hat{T} + \tilde{\mathrm{V}} \approx \hat{T} + \frac{1}{2} \sum_{m_1\alpha_1,m_2\alpha_2} \sum_{\mu_1,\mu_2} \tilde{\Phi}^{\mu_1,\mu_2}_{m_1\alpha_1 m_2\alpha_2}(\omega) \times u^{\mu_1}_{m_1\alpha_1} u^{\mu_2}_{m_2\alpha_2} + \cdots \tag{4}$$

Here $\hat{T}$ denotes the kinetic energy operator, and the quadratic model force constants

$$\tilde{\Phi}^{(2)} = \left( \frac{\partial^2 \tilde{\mathrm{V}}}{\partial u^{\mu_1}_{m_1\alpha_1} \partial u^{\mu_2}_{m_2\alpha_2}} \right)_0 \tag{5}$$

The whole *mean-field* model potential $\tilde{\mathrm{V}}$ is formally given by

$$\begin{aligned} \tilde{\mathrm{V}} = \langle \mathrm{V} - \mathrm{V}_0 \rangle &= \frac{1}{2} \sum_{m_1\alpha_1,m_2\alpha_2} \sum_{\mu_1,\mu_2} \langle \Phi^{\mu_1,\mu_2}_{m_1\alpha_1,m_2\alpha_2} \times u^{\mu_1}_{m_1\alpha_1} u^{\mu_2}_{m_2\alpha_2} \rangle \\ &+ \sum_{N=3} \frac{1}{N!} \sum_{m_1\alpha_1,\cdots,m_N\alpha_N} \sum_{\mu_1,\cdots,\mu_N} \langle \Phi^{\mu_1,\cdots,\mu_N}_{m_1\alpha_1,\cdots,m_N\alpha_N} \times u^{\mu_1}_{m_1\alpha_1} \cdots u^{\mu_N}_{m_N\alpha_N} \rangle \\ &= \sum_{N=2}^{\infty} \frac{1}{N!} \sum_{m_1\alpha_1,\cdots,m_N\alpha_N} \sum_{\mu_1,\cdots,\mu_N} \tilde{\Phi}^{\mu_1,\cdots,\mu_N}_{m_1\alpha_1,\cdots,m_N\alpha_N} \times u^{\mu_1}_{m_1\alpha_1} \cdots u^{\mu_N}_{m_N\alpha_N} \end{aligned} \tag{6}$$

In which $\langle \cdots \rangle$ represents the ensemble average of the PES according to particles distribution under thermal equilibrium, which is then re-mapped back to a Taylor expansion form like in Eq.(1). $\tilde{\Phi}$ satisfies the same symmetry of $\Phi$. The main difference between Eq.(6) and (1) is that the displacement in Eq.(1) should be infinitesimal, whereas it could be very large in Eq.(6).

TDEP works as the first-order approximation of SCP, by assuming $\tilde{\Phi}^{(2)}$ are frequency-independent and the ensemble average in Eq.(6) is carried out with a classical Newtonian canonical ensemble[(8)(9)]. Such obtained effective quadratic potential is temperature-dependent, so comes to its name. The free energy of the true system relates to that of SCP by

$$F \approx F^{\mathrm{SCP}} + \mathrm{V}_0 - \beta^{-1}\ln\langle\exp[-\beta(\mathrm{V} - \mathrm{V}_0 - \widetilde{\mathrm{V}})]\rangle_{\mathrm{SCP}} \tag{7}$$

The second term at the right-hand side of Eq.(7) should be negative according to the variational principle.

In practice, TDEP tries to seek the best $\widetilde{\Phi}^{(2)}$ by matching Eqs.(4-5) to a set of forces $\boldsymbol{F}_i$ and the corresponding displacements $\boldsymbol{u}_i$ that sampled from classical molecular dynamics (MD) simulations at a given temperature using the least squares fitting

$$\min_{\widetilde{\Phi}^{(2)}} \frac{1}{N}\sum_{i=1}^{N}\left|\boldsymbol{F}_i - \widetilde{\Phi}^{(2)} \cdot \boldsymbol{u}_i\right|^2 \tag{8}$$

Having determined $\widetilde{\Phi}^{(2)}$, the model dynamical matrix is obtained by lattice Fourier transformation

$$\widetilde{D}_{\alpha\beta}(\mu\mu';\boldsymbol{q}) = \frac{1}{\sqrt{M_\mu M_{\mu'}}}\sum_{m'}\widetilde{\Phi}^{\mu\mu'}_{m\alpha,m'\beta}e^{i\boldsymbol{q}\cdot(\boldsymbol{R}^{\mu}_{m}-\boldsymbol{R}^{\mu'}_{m'})}$$

where $M_\mu$ is the mass of particle $\mu$, $\boldsymbol{R}^{\mu}_{m}$ is the position of particle $\mu$ in the $m$-th primitive cell. Diagonalizing this dynamical matrix gives the vibrational frequency $\widetilde{\omega}$ of the model phonons in TDEP approximation

$$\widetilde{\boldsymbol{D}}(\boldsymbol{q}) = \widetilde{\omega}^2_{\boldsymbol{q}j}\tilde{\boldsymbol{e}}_{\boldsymbol{q}j}$$

in which $\tilde{\boldsymbol{e}}_{\boldsymbol{q}j}$ is the polarization vector of phonon mode $\boldsymbol{q}j$. Repeat above procedures for different temperatures so that temperature-dependent phonon frequency $\widetilde{\omega}_{\boldsymbol{q}j}(T)$ can be obtained.

A2. Path integral and full potential surface sampling

TDEP is not the only available method to realize SCP. Other similar approaches had been proposed. For example, the velocity autocorrelation function generated from classical MD simulations was used to extract the anharmonic density of state and dispersion relation of phonons[10][11]; a principal axes approximation was used to map the PES, with the vibrational free energy was then calculated by solving the Hartree-like vibrational self-consistent field equations, appended with a second-order perturbation treatment[12]; the stochastic self-consistent harmonic approximation (SSHA) directly minimizes the trial vibrational free energy Eq.(3) with respect to a set of parameters, in which the involved quantities are evaluated with importance sampling, stochastic integrals, and reweighting techniques[13]; the self-consistent *ab initio* lattice dynamics (SCAILD) method attempts to converge the phonon frequencies iteratively by resampling of the lattice displacements, whereas with the phonon polarization vectors being held fixed[14], just to mention a few.

TDEP might have the advantage of easy to implement, and high-order IFCs can be directly obtained when fitting Eq.(8)[15]. It however suffers at low temperature where a poor representation of the PES could be obtained. A modified TDEP which utilizes a stochastic sampling of the PES similar to SSHA was proposed[16]. All of these methods are preconditioned by assuming there must have well-defined undamped quasi-phonons *a priori*, which, however, cannot be generally guaranteed.

In quantum statistical mechanics, the partition function in canonical ensemble can be represented by using Trotter's decomposition and Feynman's path integral formalism as[17]

$$Z(\beta) = \mathrm{Tr}\left(e^{-\frac{\beta}{L}H}\right)^{L} = \int d\mathbf{R}_1 \cdots \int d\mathbf{R}_\mathrm{L}\, \rho(\mathbf{R}_1, \cdots, \mathbf{R}_\mathrm{L}; \beta) \tag{9}$$

where the density matrix $\rho(\mathbf{R}_1, \cdots, \mathbf{R}_\mathrm{L}; \beta) = \prod_{j=1}^{L} \rho(\mathbf{R}_j, \mathbf{R}_{j+1}; \beta/L)$, and

$$\rho(\mathbf{R}_1, \cdots, \mathbf{R}_\mathrm{L}; \beta) \propto \exp\left(-\sum_{j=1}^{L} \frac{mL}{2\beta\hbar^2}\left(\mathbf{R}_{j+1} - \mathbf{R}_j\right)^2 - \frac{\beta}{L}\sum_{j=1}^{L} \mathrm{V}(\mathbf{R}_j)\right) \tag{10}$$

in which $\mathbf{R}_j$ is the collective coordinates of all particles in the system at the $j$th imaginary-time slice (or bead) with a cyclic condition $\mathbf{R}_0 = \mathbf{R}_\mathrm{L}$. $L$ is the total number of imaginary-time slices. When $L \to \infty$, the density matrix $\rho$ in Eq.(10) provides a faithful description of the particle distribution of an equilibrium quantum system[(18)]. Therefore the path integral of Eq.(9) gives an unconstrained sampling of the full PES with high fidelity at temperature $\beta$, regardless of whether phonon can be defined or not.

To construct a model to extract the dispersion relation for lattice dynamics, a set of data $\{\mathbf{R}^f, \boldsymbol{F}^f, \boldsymbol{u}^f\}$ that contain a series of particle configurations sampled from the full PES using path integral algorithm, as well as the forces and displacement vectors associated with those configurations, are then employed to fit the IFCs, with Eq.(8) being adapted to

$$\min_{\widetilde{\Phi}_f^{(2)}} \frac{1}{N} \sum_{i=1}^{N} \left| \boldsymbol{F}_i^f - \widetilde{\Phi}_f^{(2)} \cdot \boldsymbol{u}_i^f \right|^2 \quad (11)$$

Here the superscript/subscript $f$ denotes that the quantity is for the full PES. There are three options to establish the required PIMD data set of $\{\mathbf{R}^f, \boldsymbol{F}^f, \boldsymbol{u}^f\}$: (*i*) Taking all individual beads and forces on them into account, including the spring forces due to quantum kinetic effects; (*ii*) The same as (*i*), but excluded the spring forces; (*iii*) Taking only centroids and the forces on them into account. Considering that the quantum kinetic effects should also be described by the kinetic operator in the model Hamiltonian of phonons, it is inappropriate to include them in the potential part again. This disqualifies option (*i*). As for option (*iii*), we found that the distribution of centroids is usually much narrower than the spatial region visited by individual beads, and the effective forces on centroids are non-local, resulting from an average of those acting on the individual beads that locating at different places. It is not a faithful representation of the physical PES. Therefore, in the following implementation, we choose the option (*ii*), in which the spring contribution has been discarded, and the forces are evaluated only from the full PES by $\boldsymbol{F}_i^f = -\partial V/\partial \mathbf{R}_i$. This generalization of TDEP could be termed as *Full Temperature Dependent Potential* (FTDP) to reflect this characteristic.

Other benefits to combine TDEP with path integral include that the full free energy of Eq.(7) now becomes exact and can be obtained straightforwardly, since the approximated factor $\langle \exp[-\beta(\mathrm{V} - \mathrm{V}_0 - \tilde{\mathrm{V}})] \rangle_{\mathrm{SCP}}$ is replaced by $\langle \exp[-\beta(\mathrm{V} - \mathrm{V}_0 - \tilde{\mathrm{V}})] \rangle_f$ that could be readily calculated by re-using the configurations sampled from path integral. It is impossible in the original TDEP because the employed MD data do not correspond to the real physical distribution when at low temperatures[(9)].

B. Implementation of TDEP and FTDP

To implement TDEP, first-principles simulation code such as VASP can be used to generate a classical Born-Oppenheimer MD (BOMD) trajectory that samples a canonical ensemble for a given density and temperature. Instead of using the whole data set of the MD trajectory and the corresponding forces, we prefer to extract a small number of un-correlated configurations (usually several tens) from the MD, and then recalculate the forces with a higher enough accuracy. The TDEP surface is then obtained by fitting to these refined data, rather than the raw trajectory. At first sight, it might seem this strategy is more expensive due to the extra force recalculation. However, as it turns out, this approach allows a much shorter MD simulation, and the total computational cost actually becomes less demanding. A post-processing script is utilized to automatize these procedures in our implementation.

When solving Eq.(8) or (11), a mapping table among all IFCs constructed beforehand is employed. Only independent IFCs are finally used to fit the forces. This little feature allows us to eliminate the dependence on an external commercial software[(8)] or to work with a big sparse matrix[(15)]. Other benefits of this approach include that the mapping table can be stored for later *reuse*, as long as the primitive cell and supercell keep the same setting. After having IFCs, dynamical matrix and phonon spectra are derived following the conventional routine of lattice dynamics, from which the vibrational entropy and free energy can be computed.

By far, TDEP has been mainly applied with classical MD. The shortcoming is the corresponding particle distribution becomes unphysical when the temperature is below the Debye temperature. The same problem holds for the IFCs and the resultant phonons(17). In this work, we eliminate this difficulty by implementing it associated with PIMD to achieve FTDP, in which the genuine temperature-dependent quantum distribution of nuclei is acquired. Uncorrelated configurations of atomic structure along both the real and imaginary times are picked out to refine the forces. These extracted configuration-force pairs are then fed into the standard procedure to derive the temperature dependent dynamical matrix and phonon dispersions.

C. Computational details

All HA or QHA phonons are calculated by using the frozen phonon and small-displacement method as implemented in PHON package(19). The required forces are generated by accurate first-principles calculations based on density functional theory (DFT) of electrons(20) as implemented in VASP(21)(22), in which an additional (the third one) supporting grid is used for the evaluation of the augmentation charges, for the purpose to achieve an absolute convergence in the calculated force.

Anharmonic phonons are calculated by using TDEP or FTDP scheme as outlined above. The required configurations are sampled from *ab initio* MD (AIMD) or AI-PIMD at the given temperature, and the forces are then re-evaluated with a higher precision on a par with the QHA ones.

All MD simulations are carried out in *NVT* ensemble, in which the dynamical motion of nuclei is integrated using the forces given by Hellmann-Feynman theorem, and the electrons described by DFT and the Perdew-Burke-Ernzerhof (PBE) generalized gradient approximation for exchange-correlation functional[23]. Projector augmented-wave (PAW) pseudopotential is employed for the nucleus-electron interaction[24][25]. The electronic wavefunctions are expanded using a plane-wave basis.

Specifically, for aluminum calculation, a 5×5×5 supercell with respect to the primitive FCC lattice is employed, which contains 125 atoms. The AIMD simulation is carried out at 75 K, with an energy cutoff of 400 eV for the wave-function expansion and a 4×4×4 Monkhorst-Pack (MP) mesh of k-points. In the force evaluation, a denser 6×6×6 MP mesh is used. For lithium calculation, a 5×5×5 supercell with respect to the primitive BCC lattice is employed, which has 125 atoms. The AIMD simulation is carried out at 300 K, with an energy cutoff of 500 eV and a 2×2×2 MP mesh of k-points. In the force evaluation, a denser 3×3×3 MP mesh and a higher energy cutoff of 700 eV are used.

For AIMD simulation of MH, a series of supercells up to 5×5×5 of the primitive cell of Fddd[26] and Cs-IV phase are used (the largest one contains 250 atoms), with an energy cutoff of 500 eV. An MP mesh with a size of 5×5×5 is employed to sample the k-points for the reciprocal space integration. In the force evaluation, the energy cutoff is increased to 700 eV and the k-points sampling is increased to a finer 11×11×11 MP mesh.

AI-PIMD simulations are performed to generate the nuclear quantum motion and the full PES distribution, in which up to 64 beads along the imaginary time are employed to discretely represent the Trotter decomposition of the propagator[17]. Other computational setting is the same as AIMD. The numerical convergency of all calculations was carefully checked and already been achieved with the adopted parameter setting.

## IV. RESULTS

A. Validation of the TDEP implementation

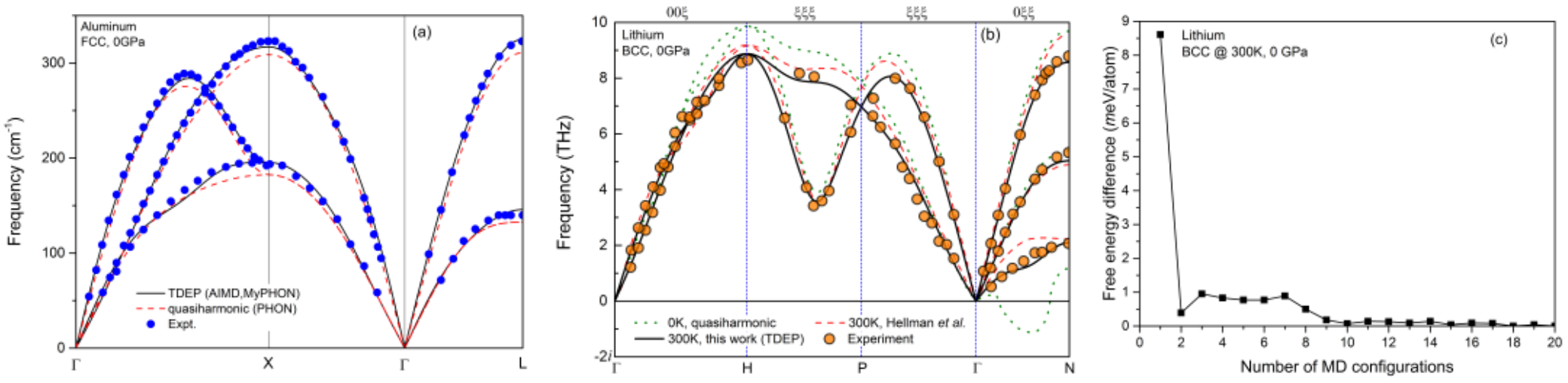

**Figure 1.** Comparison of calculated quasiharmonic and anharmonic phonon dispersion relations with the experimental data: Al (a) and Li (b) at 0 GPa, respectively. The anharmonic phonon is evaluated by TDEP with AIMD simulations at 75 K for Al and at 300 K for Li. (c) Convergence of the vibrational free energy of Li obtained from TDEP with respect to the total number of employed MD configurations.

Our implementation of the TDEP is verified by comparing the calculated phonon dispersions of Al at 0 GPa with the experimental data(27) and the QHA phonons evaluated by PHON(19), as shown in Fig.1(a). Our TDEP results (labelled as MyPHON, which is our own homemade package to implement TDEP and FTDP calculation) are in good agreement with the experimental data(27). Though Al is usually regarded as a harmonic metal, the anharmonic TDEP phonon indeed matches better with the experimental data than the QHA one, especially at around X and L point. This perfect agreement validates our implementation of the TDEP method.

Lithium provides another confirmation of the validation of our TDEP. As shown in Fig.1(b), our calculated anharmonic phonons are in good agreement with experimental data(28). In contrast, QHA calculation predicted some modes with imaginary frequency along Γ-N direction(8). Figure 1(b) also supports the strategy to simplify the TDEP by re-evaluating the accurate forces of a small set of un-correlated MD configurations. It gives better phonon dispersions than that used the whole MD configurations and the raw forces(8). The improvement is obvious in the longitudinal acoustic mode at around H, P and N point, respectively. The convergence of the calculation with respect to the total number of employed configurations (TECs) is very fast. As indicated in Fig.1(c), three TECs are enough for the vibrational free energy to converge to a level of 1 meV/atom, and ten TECs are enough to achieve a complete convergence. This feature makes it easier to carry out TDEP, since a very short MD is usually enough.

B. Generalization to FTDP

As a straightforward generalization of TDEP, FTDP expands its application range down to below Debye temperature. This is achieved by utilizing unconstrained PIMD(17), which does not require any *prior* restriction on the sampling space like that imposed in other methods(16), and the approximated effective PES is replaced by the full and physical one.

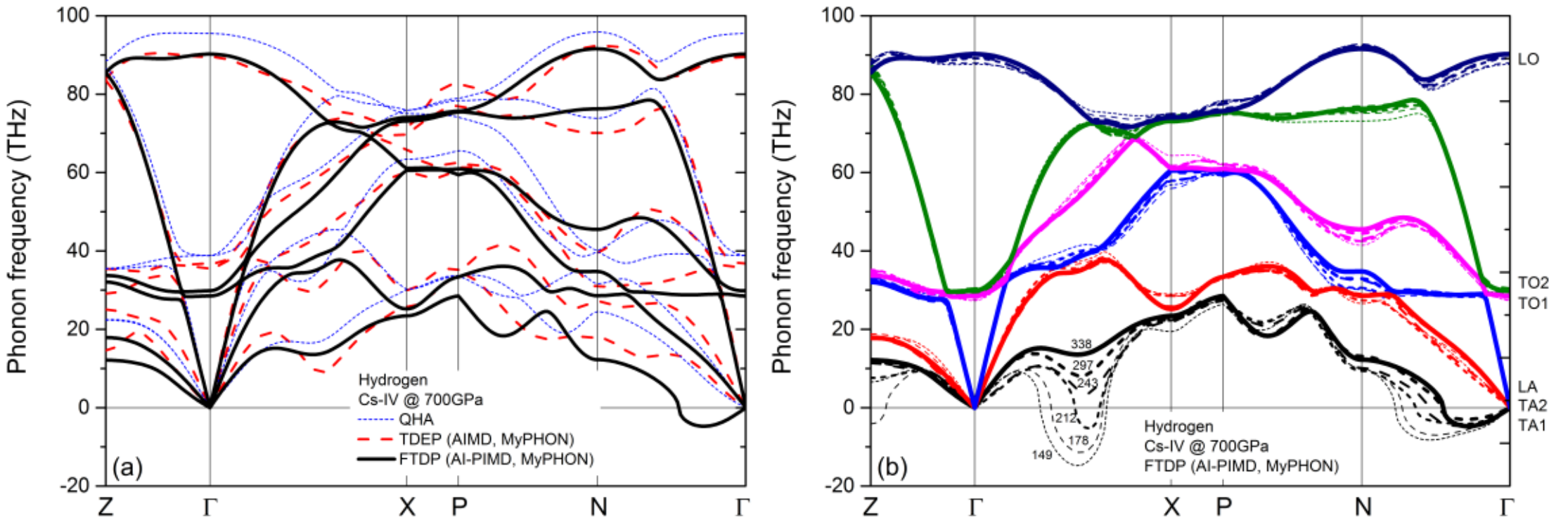

**Figure 2.** (a) Comparison of QHA and anharmonic phonon dispersion relations in Cs-IV phase of MH calculated with small displacement method, TDEP using AIMD (20 TECs) and FTDP using AI-PIMD (338 TECs) at 200 K, respectively; (b) Convergence of FTDP with respect to the total number of employed AI-PIMD configurations at 200 K. A 4×4×4 supercell containing 128 atoms is employed.

Here metallic hydrogen (MH)(29)(30) is employed to investigate the performance of FTDP and TDEP, which is assumed to be one of the most anharmonic quantum solid but without direct evidence(26)(31). Both QHA phonons of small displacement method and anharmonic phonons of TDEP and FTDP for the Cs-IV phase of MH at about 700

GPa (with an atomic volume of 0.988 $Å^3$/proton) are calculated, as shown in Fig.2(a). For TDEP and FTDP, the uncorrelated configurations are sampled from long-enough AIMD and AI-PIMD simulations performed at 200 K, respectively; and the Arabic number on each curve in Fig.2(b) denotes the used TECs for their respective FTDP computations.

It can be found that anharmonicity lowers down the overall spectra, with FTDP down more than TDEP. The main changes are (here T, L, O, A stand for transverse, longitudinal, optical and acoustical mode, respectively): (*i*) the very small gap between the TA1-TA2 mode at QHA level along $\mathrm{Z}-\Gamma$ and $\mathrm{X}-\mathrm{P}$ direction are greatly enlarged, as well as the small splitting between the TO1-LA mode at point Z; (*ii*) the anti-crossing splitting between the LO-TO2 mode is reduced to almost null; (*iii*) the small anti-crossing splitting between the TO1-TO2 at point Γ is slightly enlarged, with a large drop both of their frequencies by FTDP; (*iv*) the TO1 mode along $\mathrm{X}-\mathrm{P}$ direction is moved down greatly, and becomes close to the LA mode in FTDP; whereas the small anti-crossing splitting between the LO-TO2 mode along this line is enlarged by TDEP, but reduced to almost null again by FTDP; (*v*) the small anti-crossing gap between the LA-TO1 mode at point N is enlarged by both TDEP and FTDP; (*vi*) there is no imaginary mode at QHA and TDEP level, while they present along the $\Gamma-\mathrm{N}$ direction in FTDP. It is evident that the correction of FTDP over TDEP in Cs-IV phase of MH is tremendous, which is in line with the expectation arisen from the flat PES that was first discovered in Ref.(26). The same conclusion also holds when a bigger 5×5×5 supercell containing 250 atoms is employed.

It is worth pointing out that TDEP and FTDP change neither the crossing/anti-crossing topology nor the degeneracy of the QHA phonon band structure. This implies that the symmetry of IFCs dictated by lattice structure is preserved by anharmonicity, as usually expected. Nevertheless, the interaction among closely-lying modes due to anharmonic effects, which affect not only the diagonal terms of the dynamical matrices but also the off-diagonal terms, leads to mixing of these modes when performing the diagonalization procedure to acquire the eigenstates, re-dividing them into different branches according to irreducible representations of the symmetry at that reciprocal point, with each new band being a hybridization of the original ones(32)(33)(34). The group theory requires that bands of the same representation should not cross each other, and the resultant anti-crossing splitting of modes can be detected by inspecting the variation of their eigenvectors, as illustrated in the Supporting Information. We also note that though usually is regarded as harmonic, QHA given by small-displacement method actually contains some anharmonic effects, as evidenced by several non-negligible avoided crossing in its phonon dispersions of MH. The variation in the phonon spectra from QHA to TDEP and FTDP as shown in Fig.2, especially the enlargement or reduction of the anti-crossing gap between the modes, characterizes the variation of the anharmonic strength captured by different level of theoretical methods.

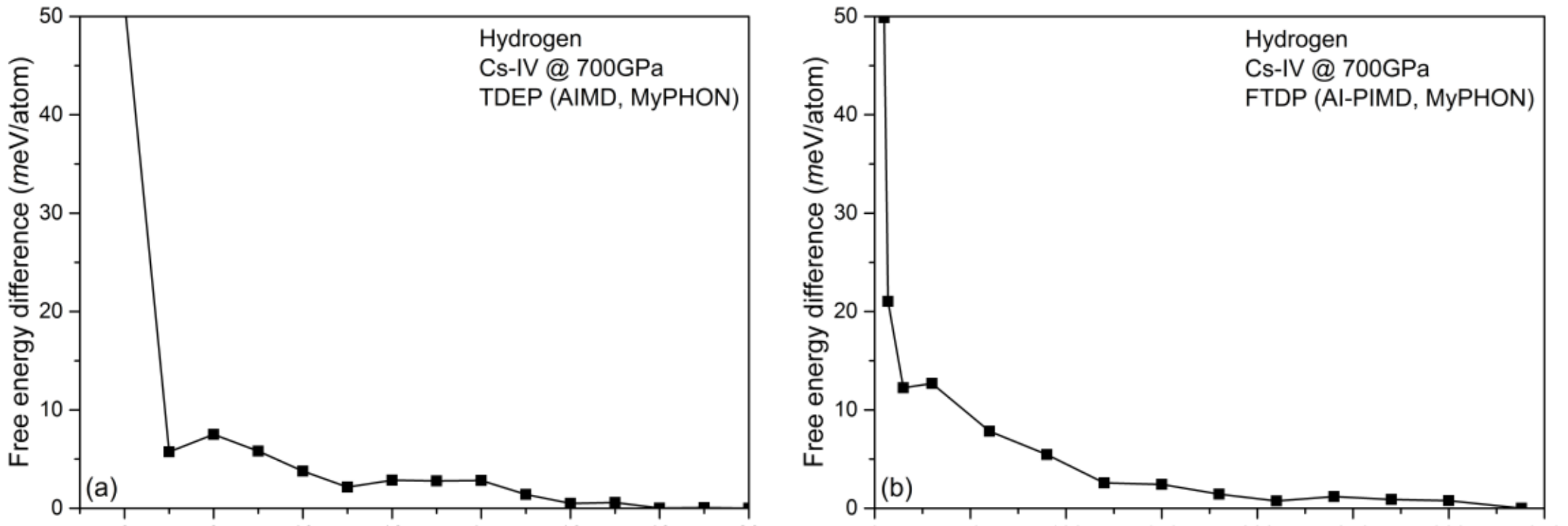

**Figure 3.** Convergence of the vibrational free energy of MH in Cs-IV phase with respect to the TECs: calculated with (a) TDEP and AIMD, (b) FTDP and AI-PIMD.

The convergence in the vibrational free energy at 200 K with respect to the TECs is shown in Fig.3. Here following the commonly adopted practice for saddle point or transition state, the modes with imaginary frequency are discarded when evaluating the vibrational free energy. It is evident that 18 TECs are enough to fully converge the TDEP, in line with the case of lithium as displayed in Fig.1. Nonetheless, it becomes much slower when in FTDP, and more than 200 TECs are required to converge the free energy to better than 1 meV/atom. In phonon dispersions, more than 300 TECs are required for a complete convergence, as shown in Fig.2. We found that it is the TA1 mode with small frequency that converges the slowest, with the most impacted region of NQE locating near the zone center (*i.e.*, at the middle between $\Gamma - \mathrm{N}$, $\Gamma - \mathrm{X}$, and $\Gamma - \mathrm{Z}$). All other modes converge much faster, and the qualitative changes in the spectral feature from QHA or TDEP to FTDP are insensitive to the TECs.

We notice that by comparison to the QHA phonons, or the TDEP phonons that sampled from a classical distribution of Newtonian trajectories, there are imaginary modes along $\Gamma - \mathrm{N}$ direction in FTDP. These small imaginary frequencies do not necessarily indicate a dynamical instability of the Cs-IV phase at the given pressure and temperature. Indeed, there had a lot of structure-searches been carried out in this region, and no new structure (*other than the Cs-IV and Fddd phases*) with lower enthalpy was found[(26)(29)(35)]. Furthermore, we also reoptimize the configurations extracted from AI-PIMD simulations of a 5×5×5 supercell. All of them relax back to the ideal lattice immediately. This implies that if there were a more stable structure, it would be very complex and have more than hundreds of atoms in its primitive cell, which is quite unlikely given the pressure is so high. Structural optimization with a disturbance according to the given vibrational mode also does not lead to a new structure. For example, for the TA1 mode at (0, 0.1, 0) point along $\Gamma - \mathrm{N}$ direction with an imaginary frequency of 3.3*i* Thz that requires a 1×10×1 supercell to accommodate, we displaced the atoms in the supercell according to the mode eigenvector with an amplitude of 0.48

Å that being comparable to the proton drift size in AI-PIMD. When relaxing this distorted structure using the Hellmann-Feynman forces, it directly restores to the original perfect Cs-IV structure (see the Supporting Information). We checked several other q-points, and the same conclusion was obtained. This result suggests that a positive frequency for many of these imaginary modes can be obtained if the employed supercell is large enough to accommodate their eigenstates.

The slow convergence and small imaginary modes aforementioned might arise from NQE and the very large amplitude of zero-point motion (ZPM) of MH at this pressure. Indeed, the AI-PIMD calculated density distribution of protons is very spreading (see Fig. 4), which even makes some proton's displacement larger than half of the nearest neighbor distance. Namely, solid MH must have a Lindemann ratio greater than 0.5. By contrast, for a classical solid, the empirical rule tells that melting should occur if the Lindemann ratio becomes greater than 0.15~0.2. Furthermore, we checked the AI-PIMD configurations with very large displacements to probe whether swapping of protons or creation of defects could be resulted. Unexpectedly, for all of these configurations, the displaced protons all relax back to their ideal position when re-optimize the structure. We also note that in AI-PIMD simulations, the nearest neighboring distance between H-H is greater than 0.9 Å, which is large enough to exclude any proton-proton quantum exchange. Namely, neither classical nor quantum particle exchange can be involved. On the other hand, we did not observe any bonds for molecules or clusters[(26)(35)]. In this regard, the imaginary modes in FTDP have nothing to do with the rotational or librational motion of molecules or clusters. Following this reasoning, we conclude that the slow convergence rate and small imaginary modes should be the consequences of the colossal NQE of MH, which leads to long-ranged interatomic interaction and reduces the reliability of Fourier interpolation when computing the phonon spectra using a supercell method. We verified this argument by carrying out a calculation of the phonon frequency on a 5×5×5 q-point grid of the Brillouin zone that is commensurate with the employed 5×5×5 supercell. All of the frequencies are positive, confirming that the imaginary modes could be spurious. We believe that a phonon spectrum without any imaginary mode could be obtainable if large enough supercell is used (as implied by the above example where a 1×10×1 supercell indeed eliminates the imaginary frequency at the (0, 0.1, 0) q-point). We estimate the required supercell size by inspecting the distribution range of the imaginary modes away from the commensurate q-points in the reciprocal space for the Fourier interpolation. The distance for a 5×5×5 supercell case is about 0.05 away if measured in the length unit of the primitive cell reciprocal vector. It thus implies that a very large supercell up to 20×20×20 might be required to get the correct Fourier interpolation and to eliminate all spurious imaginary modes. Unfortunately, we cannot confirm this inference because the required computational resource demanding is far beyond what we can afford. (Note: we cannot completely exclude the possible involvement of new lower energy structure for some of these imaginary modes, even though it seems quite unlikely based on our above analysis. The final answer for this problem can be accomplished only when one can reach a supercell as large as 20×20×20 to eliminate all of the imaginary modes or map out the whole free energy surface to display the global free energy minima, which is far beyond the scope of this work).

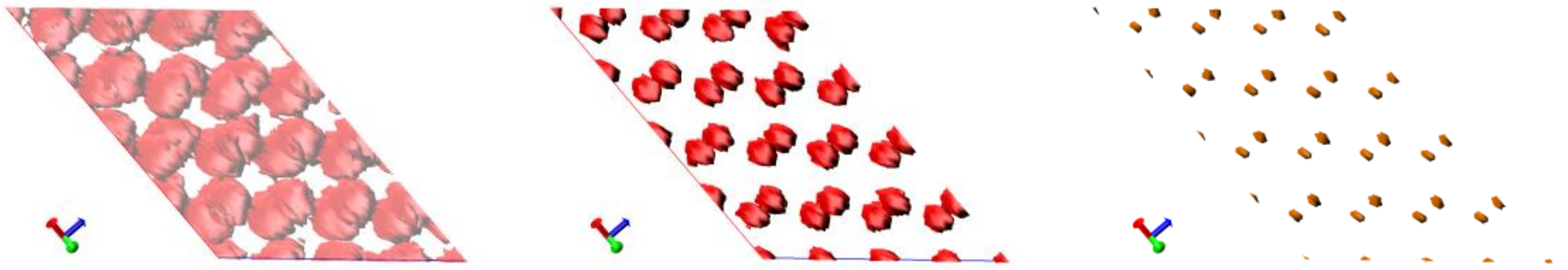

**Figure 4.** Illustration of the spatial region covered by different method (represented by averaged density distribution of protons): (left) FTDP, (middle) TDEP, and (right) QHA. Only FTDP takes the real temperature-dependent proton distribution into account. Note that in the left panel the neighboring protons visit the overlapped region only at different time. They are actually well separated at each imaginary time slice in the PIMD simulations.

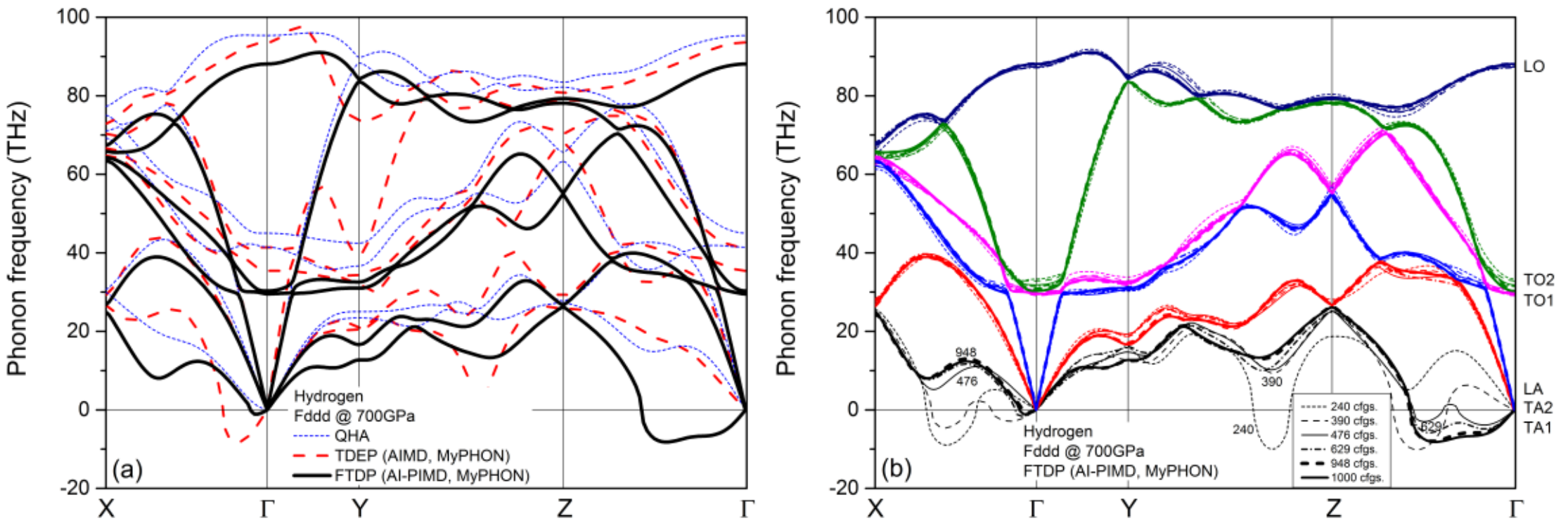

**Figure 5.** (a) Comparison of QHA and anharmonic phonon dispersions in Fddd phase of MH calculated with small displacement method, TDEP using AIMD (20 TECs), and FTDP using AI-PIMD (1000 TECs) at 200 K, respectively; (b) Convergence of FTDP with respect to the AI-PIMD TECs at 200 K. A 4×4×4 supercell containing 128 atoms is used.

The QHA and anharmonic TDEP and FTDP phonon dispersions of competing Fddd phase[26] of MH at 700 GPa calculated at different theoretical levels are shown in Fig.5(a). It is evident that at QHA level there is no imaginary mode. Contrary to usual expectations, small imaginary modes appear along $\Gamma - X$ direction at higher level of TDEP, and along both $\Gamma - Z$ and $\Gamma - X$ direction at the highest level of FTDP, respectively.

Furthermore, a large variation in the dispersion relation of the TA1 mode with respect to the level of approximation is observed. Besides the overall downwards-shift of the spectra from QHA to TDEP and FTDP, as well as some enlargement or reduction of the anti-crossing gap between coupled modes due to anharmonicity, we find that LO and TO2 modes of TDEP dip deeply at point Y, showing a striking contrast with respect to QHA and FTDP. The convergence rate of the phonon spectra of Fddd with respect to TECs is much slower than for Cs-IV phase, and more than 900 TECs are required for a complete convergence of the TA1 mode, as indicated in Fig.5(b). Similar to Cs-IV phase, however, other branches converge very fast.

C. Comparison in thermodynamic property

Above results demonstrate that when the theoretical level of approximation is increased continually along the ladder from QHA and TDEP to FTDP, distinct anharmonic features in the computed phonon dispersions can be observed. It is interesting to check whether these deviations lead to discernible change in the vibrational free energy or not.

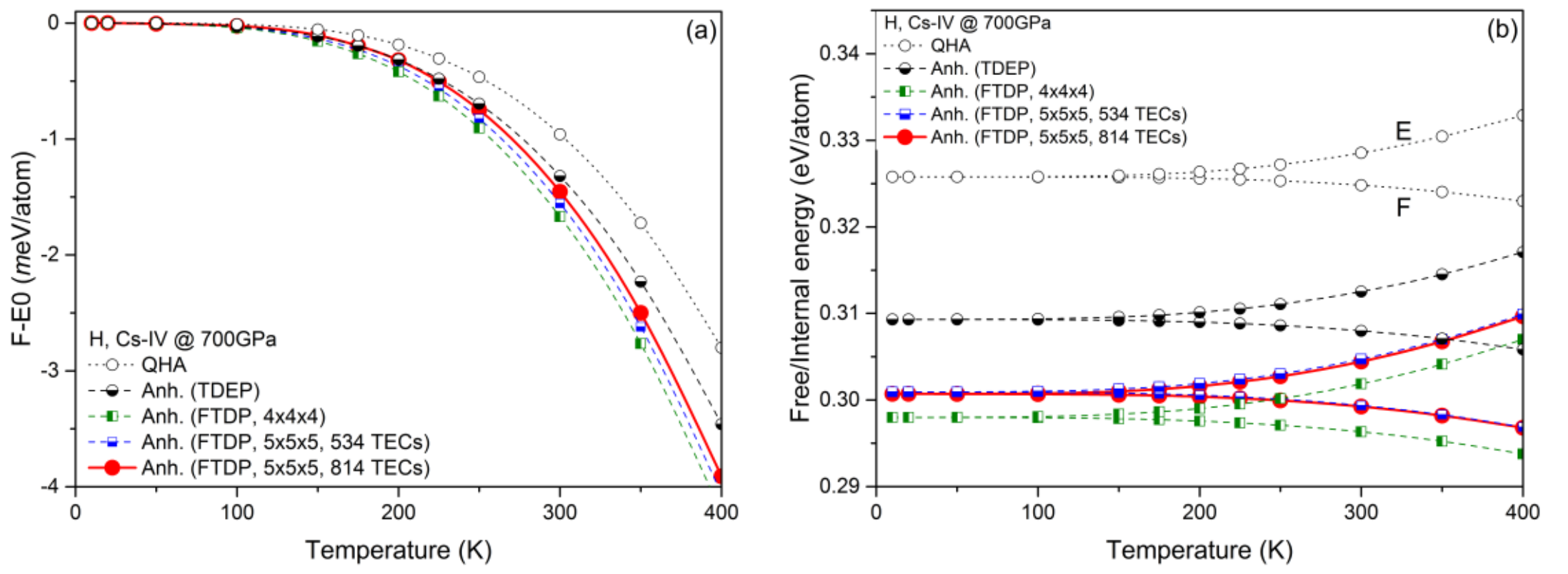

**Figure 6.** Variation of the vibrational energy of MH in Cs-IV phase as a function of temperature calculated with different methods. (a) the relative free energy with respect to ZPE, (b) the total vibrational free energy (F, lower branch) and internal energy (E, upper branch). The ZPE obtained with a 5×5×5 supercell and 534 TECs is within 0.08% of that with 814 TECs.

Figure 6(a) displays the relative variation of vibrational free energy of MH in Cs-IV phase at 700 GPa as a function of temperature calculated by different methods or parameter setting. It is evident that the anharmonicity is significant: at 300 K, the contribution to the relative free energy from QHA is -0.96 meV/atom, whereas it decreases to -1.32 meV/atom when goes to TDEP, with a correction of 37.5%; and the further correction to TDEP from FTDP is about 26.5%. On the other hand, increase the employed supercell size of FTDP to 5×5×5 leads to an additional correction about - 6.9%, with the converged result of -1.45 meV/atom being obtained with 814 TECs.

The total vibrational free energy was compared to the internal energy as illustrated in Fig. 6(b). One of the implications is that the entropy contribution is less than 0.5% and 1.3% when at 300 K and 400 K, and is negligible when below 150 K, respectively. The relative shift in ZPE between different methods, however, could reach a magnitude as large as 9%, with a correction to QHA from TDEP is about 5%, and the correction to TDEP from FTDP is less than 4%. On the other hand, increase the supercell size to 5×5×5 gives an additional correction less than 1%, for which using 534 or 814 TECs leads to a difference less than 0.07%. This indicates the convergence in the total free/internal energy. It also can be inferred that if larger supercell beyond 5×5×5 is employed to eliminate the imaginary modes, the resultant correction should be less than 1%.

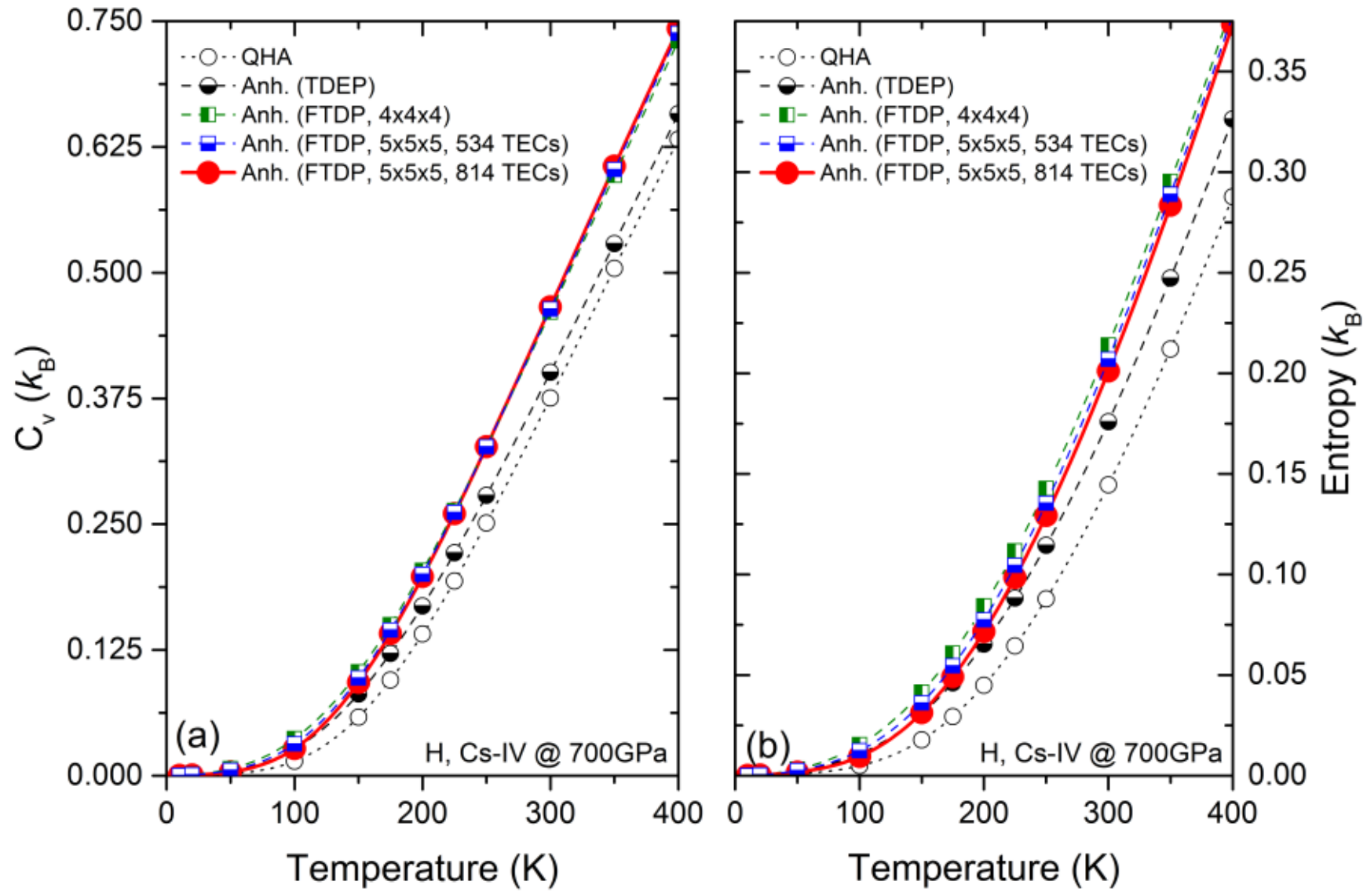

**Figure 7.** Variation of (a) vibrational specific heat and (b) entropy of MH in Cs-IV phase as a function of temperature obtained with different methods.

Similar conclusion holds for vibrational entropy and specific heat, in which only QHA and TDEP manifest a perceptible deviation, as shown in Fig.7. For example, the deviation of entropy from the converged one at 300 K is about 13% and 28% for TDEP and QHA, respectively. This magnitude is much greater than the difference between FTDP results with different parameter setting. It is thus verified that TDEP is inadequate for lattice dynamics of solid MH.

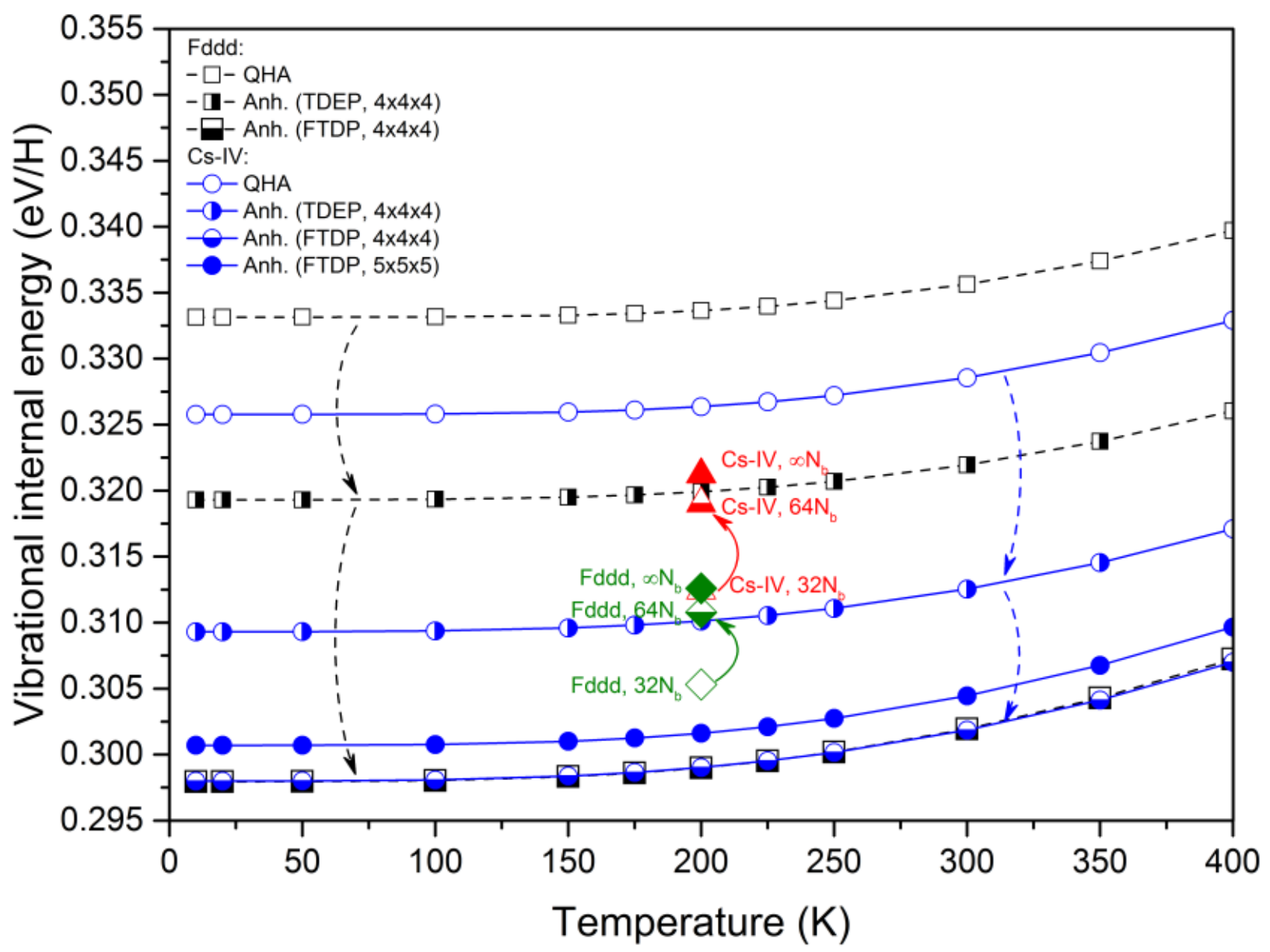

**Figure 8.** Comparison of the vibrational internal energy calculated by different methods for MH at 700 GPa as a function of temperature. For direct AI-PIMD simulations, the (open, half-filled, and filled) up-triangle points indicate that of Cs-IV phase calculated using a 5×5×5 supercell, and those of rhombus points are for that of Fddd phase using a 4×4×4 supercell; The employed number of beads ($N_b$) is also marked, which was extrapolated to the infinite limit by using the Richardson formula[36].

Aforementioned results indicated that the vibrational properties of phonon or quasiphonon are *converging* from QHA to TDEP and then to FTDP, and it is converged when in the latter. However, since all of these methods inherit an implicit approximation of to map the results onto a functional framework of harmonic oscillators, it is possible that such constraint could drive the "converged" results away from the physical one. For that purpose, benchmark verification is carried out by comparing the calculated internal energy with that of direct AI-PIMD simulations.

The final results are shown in Fig.8. It is surprising to observe that within the phonon framework the correction from anharmonicity is to decrease the vibrational internal energy, with the same trend for both TDEP and FTDP, and was observed in both Fddd and Cs-IV phase of MH, as indicated by the dashed arrows in the figure.

On the other hand, direct AI-PIMD simulations give a much higher internal energy, which becomes more higher when extrapolated to the thermodynamic limit, as the solid arrows in Fig.8 indicated. Specifically, with respect to the "converged" results of FTDP, the vibrational internal energy given by direct AI-PIMD simulations at 200 K is about 20 meV/H (14 meV/H) higher, accounting for a deviation of 6.6% (4.5%), for the Cs-IV (Fddd) phase, respectively. Here, the most significant feature is the *opposite* trend of the corrections in the hierarchical phonon framework (*i.e.*, from QHA, TDEP, and to FTDP) against the direct AI-PIMD simulations (as well as their failure in prediction of the relative order of the vibrational internal energy between the competing Cs-IV and Fddd phase). We estimated that using a larger supercell in FTDP is unlikely to remedy the discrepancy. This deteriorated performance of phonon method should be a consequence of forcibly mapping the complex PES back to a quadratic form of phonons. Figure 8 thus is a strong indicator that phonon or quasi-phonon formalism might have broken in solid MH.

## V. DISCUSSION

### A. On the Limitation of FTDP and SCP

The total energy in SCP may not equal to the sum of the band energy of quasi-phonons. Their difference can be measured by

$$\Delta E \approx \langle \mathrm{V} - \mathrm{V}_0 - \frac{1}{2}\sum_{m_1\alpha_1,m_2\alpha_2}\sum_{\mu_1,\mu_2}\tilde{\Phi}^{\mu_1,\mu_2}_{m_1\alpha_1 m_2\alpha_2} \times u^{\mu_1}_{m_1\alpha_1} u^{\mu_2}_{m_2\alpha_2}\rangle_{\text{quasi-phonon}} \quad (12)$$

This approximate evaluation is restricted in the space spanned by the quasi-phonons. The compensated contribution from surrounding medium should be small if quasi-phonon as independent quasi-particle is valid. An exact evaluation of $\Delta E$ without any constraint can be obtained in FTDP with the help of path integral

$$\Delta E = \langle \mathrm{V} + \hat{E}_k - \mathrm{V}_0 - \frac{1}{2}\sum_{m_1\alpha_1,m_2\alpha_2}\sum_{\mu_1,\mu_2}\tilde{\Phi}^{\mu_1,\mu_2}_{m_1\alpha_1 m_2\alpha_2} \times u^{\mu_1}_{m_1\alpha_1} u^{\mu_2}_{m_2\alpha_2} - \hat{E}_k\rangle_{\mathrm{PI}} = \langle E - \mathrm{V}_0\rangle_{\mathrm{PI}} - E^{\mathrm{FTDP}}_{\text{quasi-phonon}} \quad (13)$$

in which the kinetic energy contribution is explicitly included. The departure of $\Delta E$ from zero signals how far the system is away from the presumed undamped quasi-phonon model. It can be seen from Fig.8 that $\Delta E = 10{\sim}20$ meV/H for MH, which is large enough to invalidate the picture of phonon or quasi-phonon in solid MH. To our best knowledge, it is the first report of this kind of breakdown, and comparison between FTDP and direct PIMD simulations is an effective strategy to probe this anharmonic phonon failure.

From Eqs. (1) and (6), it is evident that $\tilde{\Phi}^{\mu_1,\cdots,\mu_N}_{m_1\alpha_1,\cdots,m_N\alpha_N} \neq \Phi^{\mu_1,\cdots,\mu_N}_{m_1\alpha_1,\cdots,m_N\alpha_N}$. The former can be viewed as an ensemble-averaged representation of the latter, and is temperature-dependent. It should be noted that generally the mapping of the PES back to a Taylor expansion series as shown in Eq.(6) may not be unique if the displacement vectors are not infinitesimal. To achieve a unique mapping as close as possible, however, a recursive header-approximation that pushing all contributions into the lowest-order terms could be employed, for which the *i*-th order IFC satisfies $\tilde{\Phi}^{(i)} \equiv \tilde{\Phi}^{(i)}|_{\max_{j<i}\{\Phi^{(j)}\}}$. In this sense, the TDEP/FTDP, as well as other non-perturbative SCP methods[(7)(11)(12)(13)(14)], all can be regarded as a kind of header-approximation, in which the quadratic order contributions are maximized.

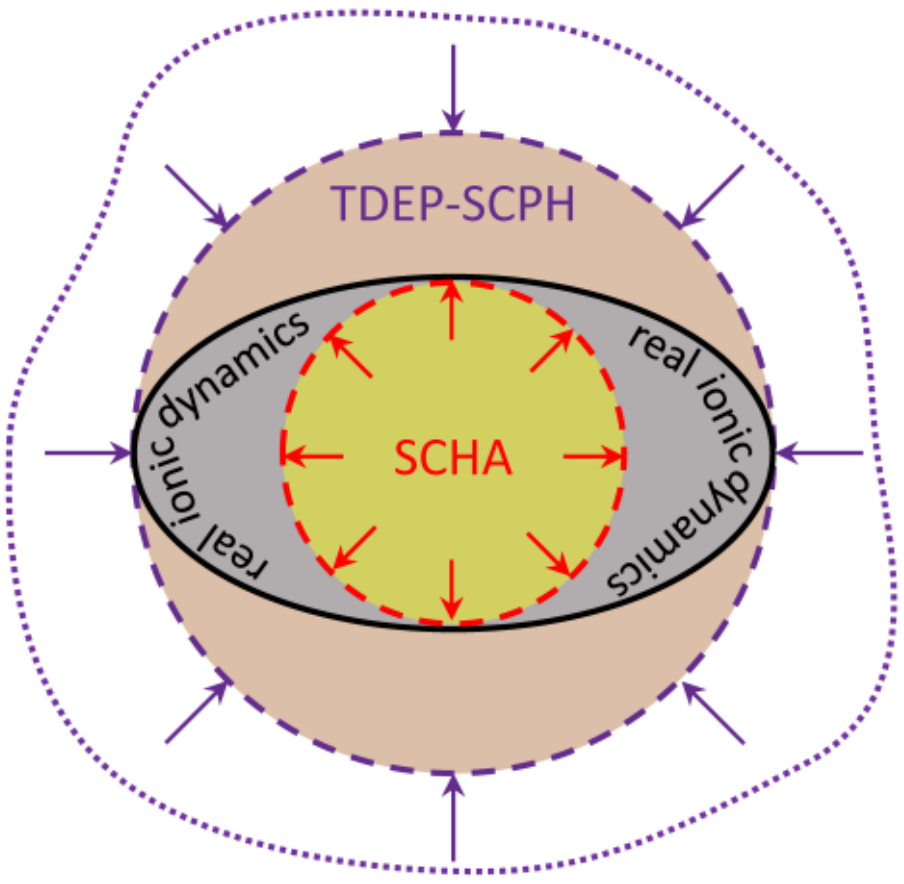

**Figure 9.** Schematic illustration of two different types of strategy to converge towards the real ionic dynamics. The first type starts from harmonic approximation such as SCHA, and the sampling space is constrained by QHA formalism; it approaches the real ionic dynamics from inside by minimization of the QHA free energy. The second type (such as TDEP or FTDP) lifts all constraints at first, and then approaches the real ionic dynamics from outside by mapping the PES back to a quadratic form. The two should be identical if the system is harmonic or nearly harmonic, but could be different for a general case.

Generally, there are two types of strategy to extract the header contribution from the PES, but with a quite different philosophy behind them. The first one starts by assuming that a good harmonic effective PES always can be achieved, and the corresponding phonons are employed *in advance* (with undetermined parameters) to construct the SCP formalism, then a series of operations are performed to optimize the phonons. In this approach, the ensemble-averaged $\tilde{\Phi}$ is optimized in a sampling space spanned by the effective harmonic phonons that minimize the residual PES difference (or the free energy). Most SCP methods belong to this category. The second type (like TDEP or FTDP) does not assume an effective harmonic PES *a priori*. The ensemble-averaged $\tilde{\Phi}$ is performed over a space sampled by un-constrained MD or PIMD simulations. The quasi-phonons are then derived from the fitted PES. The relationship of these two strategies is schematically illustrated in Fig. 9. It is evident that when the PES can be well represented by "averaged" phonons, identical results should be obtained. In this special situation, TDEP/FTDP restores an effective "harmonic" PES, and all higher-order $\widetilde{\Phi}^{(i)}$ for $i \geq 3$ should vanish. On the other hand, if the high-order terms are small enough, a perturbation treatment on top of SCP might be applied. If this re-renormalized expansion is still non-convergent, however, the phonon concept cannot be implemented and may break completely.

**TABLE 1**. Hierarchy of theoretical approaches for lattice dynamics

| Expansion level | | PES and perturbation | | Non-perturbative method: Effective PES mapping | | Direct method (full quantum mechanics) |
|---|---|---|---|---|---|---|
| | | HA/QHA | Anharm. | SCHA | TDEP/FTDP | |
| $1^{st}$ order | | -- | -- | Equilibrium position correction (harmonic NQE) | Equilibrium position correction | (1) solve ionic Schrödinger equation directly<br>(2) perform path integral simulation |
| $2^{nd}$ order | | undamped phonon | -- | undamped quasi-phonon (renormalized phonon) | undamped quasi-phonon (renormalized phonon) | |
| perturbations | $3^{rd}$ order | -- | Three-phonon scattering | Can be included via higher order cumulant expansions | Three-quasiphonon scattering | |
| | $4^{th}$ order | -- | Four-phonon scattering | | Four-quasiphonon scattering | |

The theoretical hierarchies for anharmonic lattice dynamics (for all currently available and those can be expected in the near future) are listed in Table 1. The traditional HA/QHA and perturbation method on top of HA is appropriate for weakly anharmonic system. For moderately strong anharmonicity, non-perturbative methods such as SCHA or TDEP/FTDP are required. TDEP can be applied only when the temperature is much higher than the Debye temperature, *i.e.*, $T \gg \theta_D$; whereas FTDP is required if $T < \theta_D$. In extreme situation where the lattice dynamics is so violent that quasi-phonon as independent identity of physics invalid, FTDP might fail, and one has to resort to the full quantum ionic Schrödinger equation or path integral simulations.

The possible failure of phonon or quasi-phonon as exhibited in MH is traced to the large deviation of the eigenstate and wavefunction of quantum collective motion of nuclei far away from the solution of a quadratic model: *it deteriorates, rather than improves, the result when using a quadratic form to model the local PES*. On the other hand, any PES around an equilibrium position in principle should be approximated by a quadratic form, as long as the associated region is small enough. In this regard, the possible failure of quasi-phonon in MH is also related to the very flat PES(26) and the exceptionally wide spatial area experienced by the ionic dynamics of protons (see Fig. 4).

B. Non-Perturbative Lattice Dynamics beyond Phonon Picture

As discussed above, the quantum eigenstate of lattice dynamics could be far away from that of an oscillatory system (*i.e.*, phonon or quasi-phonon) if the PES is complex, which makes the attempt to locally map the PES back to a quadratic form become in vain. “Direct PIMD”+“thermodynamic integral” simulation(17) provides a vital method to calculate the thermodynamics of such a strongly anharmonic system. If the dispersion relation (or, the ionic energy band) is required, however, the complex PES could be interpolated or fitted to a generic form to derive the eigenstates of the quantum ionic motion beyond the phonon picture non-perturbatively.

Since this topic is far beyond the main scope of this work, here we just sketch a theoretical formalism succinctly to illustrate what it might work out following this line, which might be helpful and inspirational for other people. Our derivation is based on the observation that the eigenstates of any quadratic PES form a complete and orthogonal basis set, so that any well-behaved function defined on the lattice can be uniquely expanded by using this basis set without loss of any generality. Therefore, the general ionic wavefunction can be expressed as an expansion of

$$\Psi_j = \sum_i c_{ij}\varphi_i \tag{14}$$

where $\varphi_i$ is the expansion basis and $c_{ij}$ are the corresponding coefficients. The secular equation for the eigenstate $\Psi_j$ of the ionic Hamiltonian is thus changed to

$$H\Psi_j = [H_0 + (V - V_2)]\Psi_j = \sum_i c_{ij} E^0{}_i \varphi_i + \sum_i c_{ij} (V - V_2)\varphi_i = E_j \Psi_j \tag{15}$$

in which $H_0 = \hat{T} + V_2$ and satisfies the secular equation of $H_0\varphi_i = E^0{}_i\varphi_i$.

Replacing the full potential $V$ in Eq.(15) with the FTDP version $V_{\mathrm{FTDP}}$, multiplying $\Psi_i^*$ from the left side and then integral over the whole space, we obtain

$$E_j\delta_{jk} = \sum_i c_{ik}^* c_{ij} E^0{}_i + \sum_{i'i} \int c_{i'k}^* (V_{\mathrm{FTDP}} - V_2) c_{ij} \varphi_{i'}^* \varphi_i dr \tag{16}$$

Here the ortho-normalization conditions of $\int \Psi_i^*\Psi_j dr = \delta_{ij}$ and $\int \varphi_i^*\varphi_j dr = \delta_{ij}$ have been used.

Diagonalization of the matrix at the right hand side of Eq.(16) gives the eigenstates and eigenenergies of the general lattice dynamics that beyond phonon picture. This approach is somewhat analogous to the widely applied pseudopotential plus planewave method for electronic subsystem, with $V_{\mathrm{FTDP}} - V_2$ corresponds to the pseudopotential and $\varphi_i$ corresponds to the planewave basis. It is necessary to note that the introduction of a reference potential $V_2$ is only for the purpose to facilitate the computation, which could be quadratic of any form. A reference potential other than quadratic also can be used, as long as the corresponding Hamiltonian $H_0$ can be solved.

The intrinsic difference of formalism Eqs.(14-16) from the SCP is obvious. The latter restricts the solution to the vicinity of a quadratic potential only. By contrast, Eqs.(14-16) represent the most general treatment of the lattice dynamics, and will reduce to SCP when the PES is close to being quadratic *averagely* [taking $c_{ij} = \delta_{ij}$ and minimizing the second term at the right hand side of Eq.(16) with respect to the variation of $\varphi_i$], and QHA is restored when PES is quadratic [when the second term at the right hand side of Eq.(16) vanishes].

The first correction of Eq.(16) upon SCP is via two-band mixing. Assume $\varphi_1$ and $\varphi_2$ are two eigenstates of SCP approximation, they form a $2 \times 2$ symmetric block in the matrix of Eq.(16). Diagonalization of this sub-matrix leads to a general solution of $\Psi_1 = \frac{1}{\sqrt{a^2+b^2}}(a\varphi_1 + b\varphi_2)$ and $\Psi_2 = \frac{1}{\sqrt{a^2+b^2}}(b\varphi_1 - a\varphi_2)$, which mixes the SCP bands into two new eigenstates, with the mixing parameters of ($a$, $b$) being determined by the submatrix elements. The eigenstate energies are accordingly modified by Eq.(16). Corrections from multi-band mixing will reshape the band structure of the ionic dispersion further, and may lead to completely unprecedented phenomenon in lattice dynamics.

It should be noted that this generalization of ionic dynamics to Eqs.(14-16) is applicable not only to solid, but also to liquid. In the latter case, replacing $\varphi$ by a plane wave basis might be recommended, which describes the collective motion of a quantum fluid of ions.

The advantage of to use FTDP in this general formalism is that $V_{\text{FTDP}}$ can be determined in advance by AI-PIMD, and the Hamiltonian is then completely fixed. It hence eliminates the requirement for additional iterations or self-consistence procedures, and avoids convergence problem in algorithm implementation. It also does not have any constraints on the shape of the PES and the ionic wave-functions, making it suitable for any ionic motions. The theoretical development along this line will drive the paradigm of lattice dynamics to shift beyond the conventional phonon picture and towards *general ionic dynamics* (GID), which is requisite in exploring unknown quantum dynamics of solid or liquid[(37)].

## VI. CONCLUSION

In summary, the TDEP for anharmonic phonon was generalized to the full potential case, *i.e.*, FTDP, in which unconstrained AI-PIMD simulation is utilized to generate the required uncorrelated configurations for PES sampling. The new method was applied to MH at high pressure. It is found that the strong anharmonicity in MH significantly slows down the convergence rate with respect to the TECs and the supercell size. The improvement of the theory from QHA and TDEP to FTDP was quantified in terms of vibrational energy and entropy. Prominent correction from FTDP was obtained.

Using direct AI-PIMD simulation as the benchmark, we further discovered the enlarged discrepancy between the most accurate SCP formalism and the exact theoretical lattice dynamics, which not only defies QHA, but also the anharmonic TDEP and FTDP, and led to the discovery of the possible breakdown of phonon picture in solid MH, illustrating the striking consequence of colossal NQE in this exotic state of matter. It suggests MH as a very strongly anharmonic quantum solid that might possess unknown properties beyond the phonon concept. A general formalism for quantum lattice dynamics was then proposed. Its reduction to the phonon approximation of HA and SCP was discussed. The capability to describe any shape of PES, even a liquid state, was highlighted. This advancement established FTDP and PIMD as an effective approach to study the moderate anharmonicity, as well as to probe the possible failure of phonon picture in strongly anharmonic system.

## Supporting Information Available

Fine details of the phonon spectra to demonstrate avoided crossing of closely-lying phonon modes, the resultant change in vibrational eigenvectors due to avoided crossing, the configurational representative of one TA mode with imaginary frequency and its spontaneous relaxation back to the original perfect Cs-IV lattice.

## ACKNOWLEDGMENTS

This work was supported by National Key R&D Program of China under Grant No. 2021YFB3802300, the National Natural Science Foundation of China under Grant No. 11672274 and the NSAF under Grant No. U1730248. Part of the computation was performed using the supercomputer at the Center for Computational Materials Science (CCMS) of the Institute for Materials Research (IMR) at Tohoku University, Japan.

**Competing Financial Interests**

The author declare no competing financial interests.

**Data Availability Statements**

The data that supports the findings of this study are available within the article.

Table of Contents (TOC) graphic

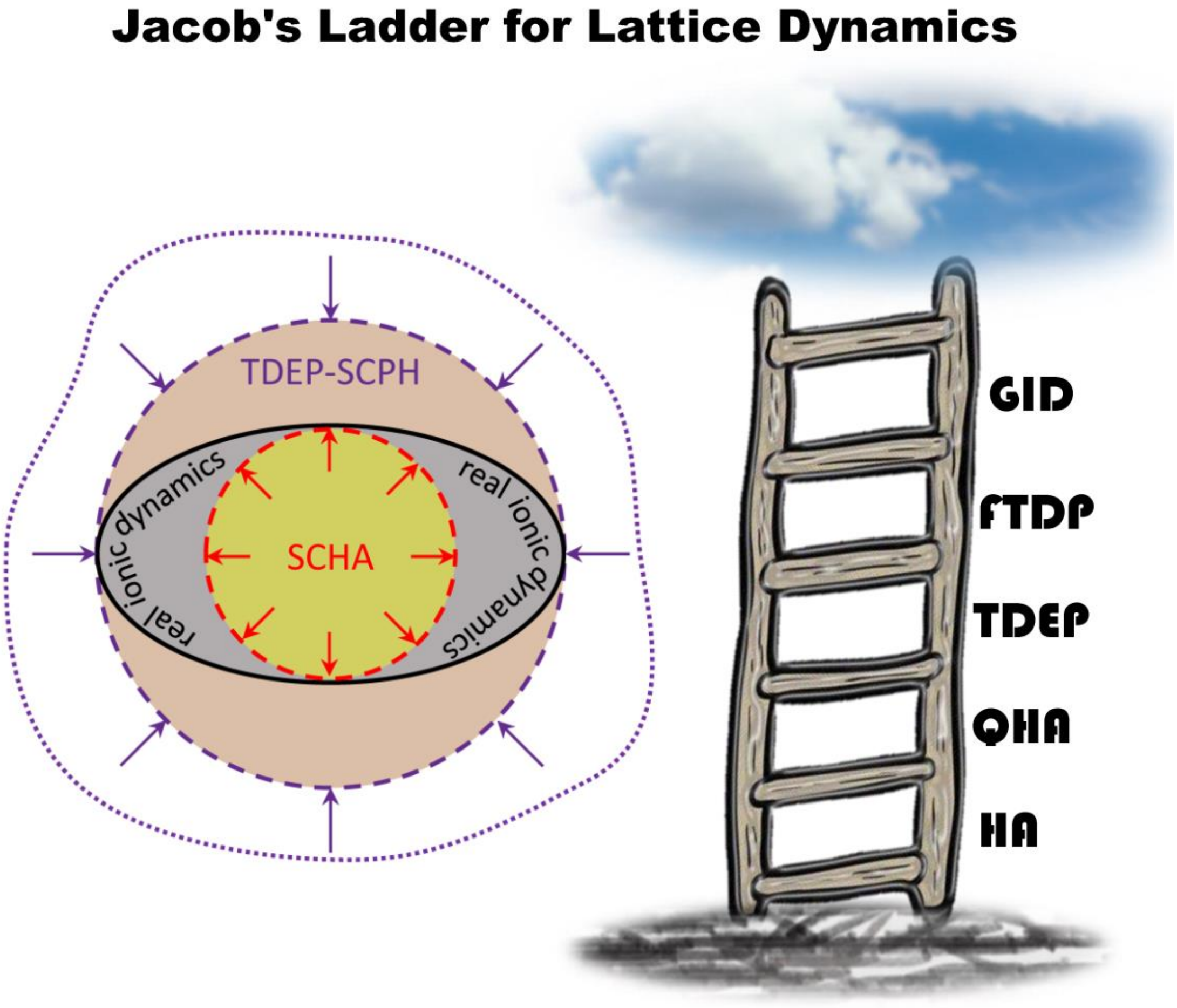

# Supporting Information for "Full Temperature-Dependent Potential and Anharmonicity in Metallic Hydrogen: Colossal NQE and the Consequences"

Hua Y. Geng[1,2†]

[1]*National Key Laboratory of Shock Wave and Detonation Physics, Institute of Fluid Physics, CAEP; P.O.Box 919-102 Mianyang, Sichuan, P. R. China, 621900*

[2]*HEDPS, Center for Applied Physics and Technology, and College of Engineering, Peking University, Beijing 100871, People's Republic of China*

† Email: s102genghy@caep.cn

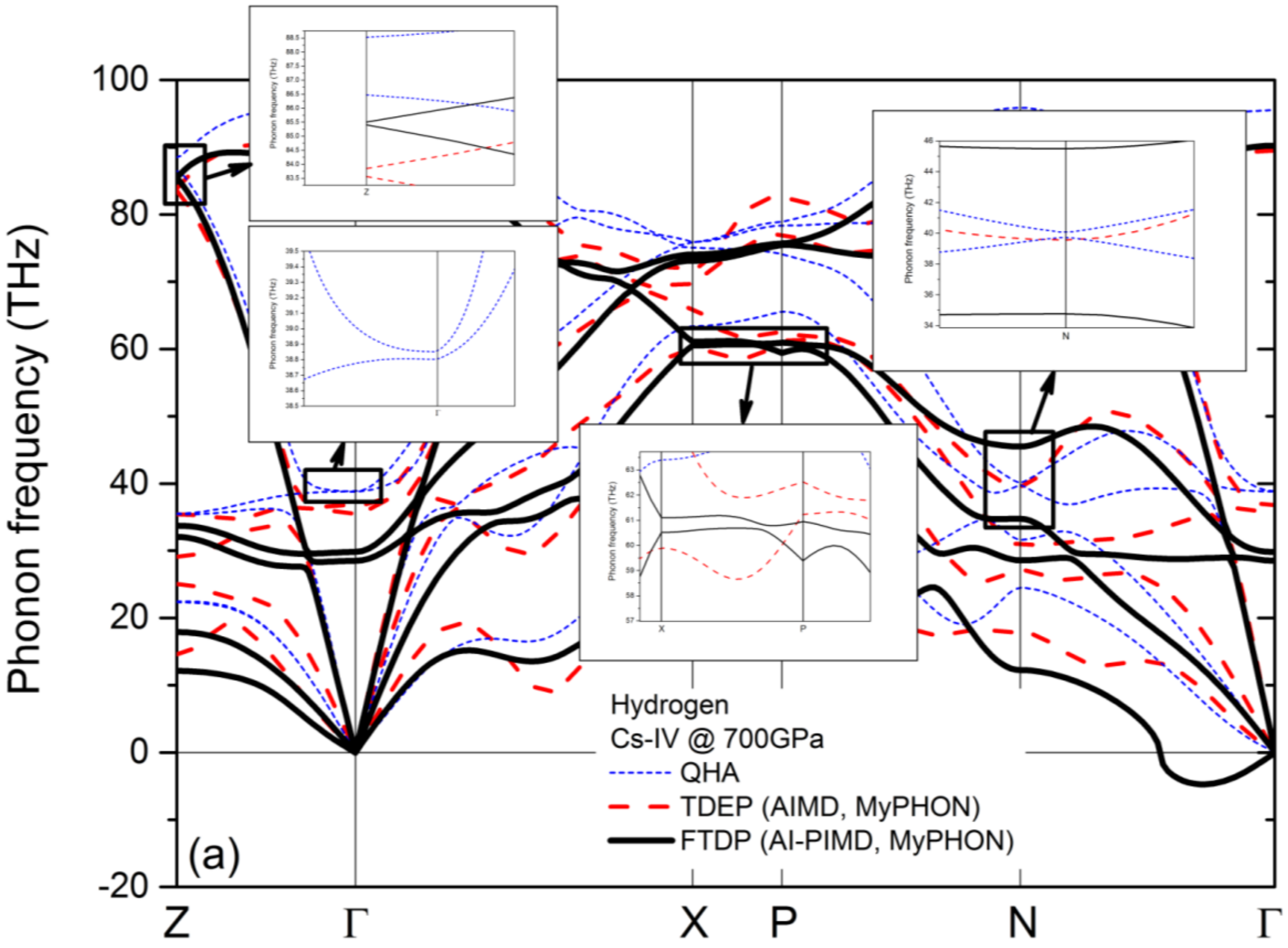

**Figure S1.** Calculated phonon dispersions of CS-IV phase of MH with a 4×4×4 supercell. The enlarged insets illustrate the typical avoided crossings of phonon normal modes due to mixing. Note: In lattice dynamics, the symmetry of lattice and phonon eigenstates dictates which modes can be mixed. On the other hand, the group theory requires that the bands belong to the same representation of the symmetry should not cross, thus leading to anti-crossing with a splitting gap in phonon band structure. The size of this avoided crossing gap is governed by the off-diagonal terms of the dynamical matrices for these coupled modes, which in turn is affected by the strength of anharmonicity that is partially described by TDEP or FTDP method.

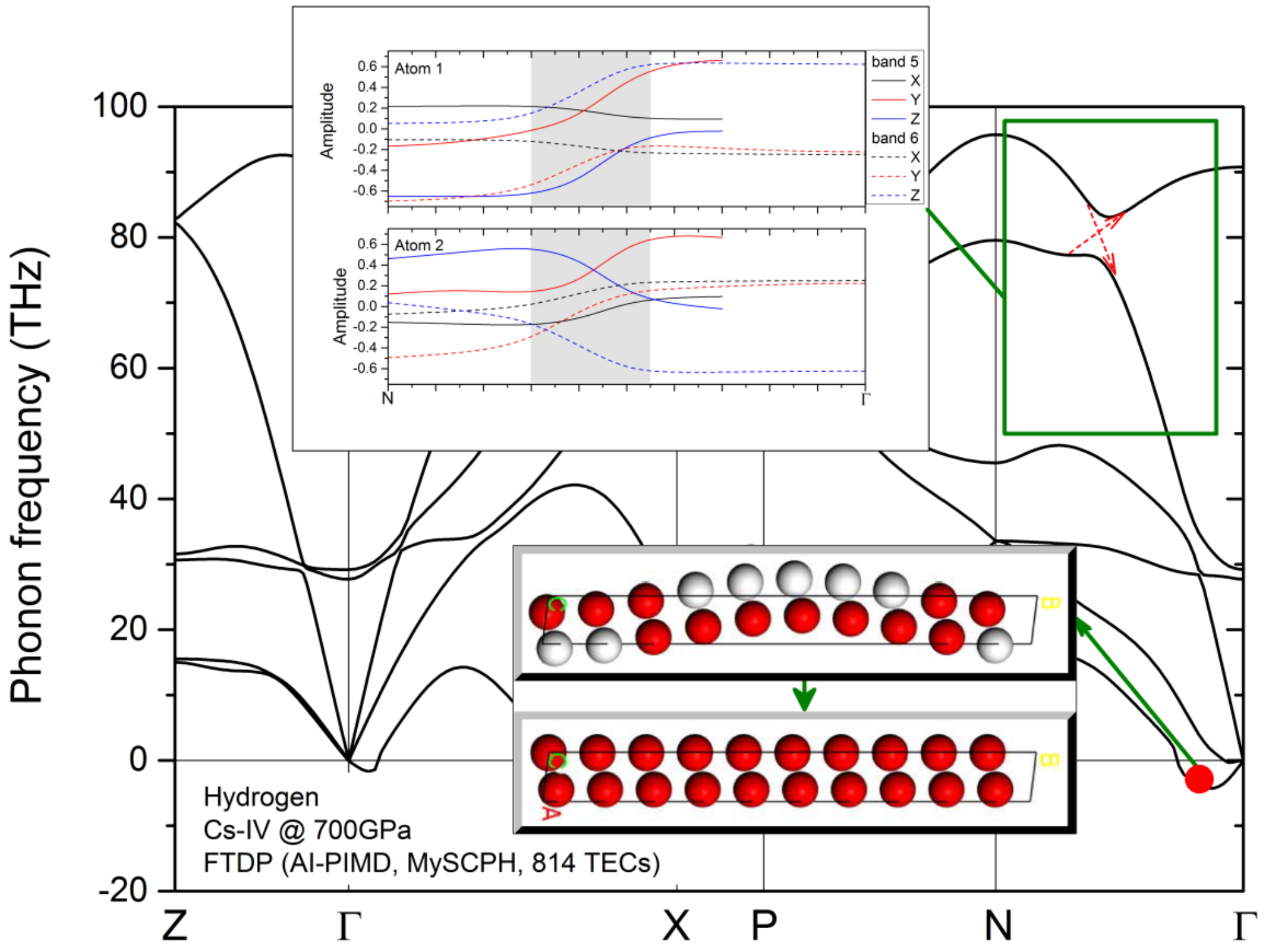

**Figure S2.** Calculated phonon dispersions of CS-IV phase of MH with a 5×5×5 supercell, in which the upper inset illustrates the abrupt change in the eigenvectors of the band 5 and band 6 due to anharmonic interaction and the resultant avoided crossing at near the "intersection point" (as the red-dashed arrows in the main figure indicated, which corresponds to the shaded region in the inset); in the lower inset the relaxation of a vibrational mode (represented within a 1×10×1 supercell) with an imaginary frequency (as the red point in the main figure indicated) back to the original perfect Cs-IV configuration (as the green arrow in the inset indicated) is highlighted. Note: By inspecting the distribution range of these imaginary modes around the supercell reciprocal lattice points represented in the primitive cell reciprocal lattice (in the case of a 5×5×5 supercell, they are only 0.05 away if measured in the length unit of the primitive cell reciprocal vector), it is estimated that a very large supercell up to 20×20×20 might be required for the purpose to remove all of these spurious modes that arising from Fourier interpolation completely.